\documentclass{ckm}                 

\usepackage{txfonts}            

\confname{Workshop on the CKM Unitarity Triangle, IPPP Durham, April
  2003}

\usepackage{epsfig}

\usepackage{rotate}

\newcommand{\lsim}{
\mathrel{\hbox{\rlap{\hbox{\lower4pt\hbox{$\sim$}}}\hbox{$<$}}}}

\newcommand{\gsim}{
\mathrel{\hbox{\rlap{\hbox{\lower4pt\hbox{$\sim$}}}\hbox{$>$}}}}

\title{Flavour Symmetry and Clean Strategies to Extract $\gamma$}

\author{R Fleischer}

\address{Theory Division, CERN, CH-1211 Geneva 23, Switzerland}

\begin{document}


\begin{titlepage}

\begin{flushright}
CERN-TH/2003-141\\
hep-ph/0306270
\end{flushright}

\vspace{2cm}
\begin{center}
\Large\bf Flavour Symmetry and Clean Strategies to Extract $\gamma$
\end{center}

\vspace{1.9cm}
\begin{center}
{\large Robert Fleischer}\\[0.1cm]
{\sl Theory Division, CERN, CH-1211 Geneva 23, Switzerland}
\end{center}

\vspace{1.7cm}

\begin{center}
{\large {\bf Abstract}}
\end{center}

\vspace{0.3cm}

\begin{quotation}
\noindent
One of the key elements in the testing of the 
Standard-Model description of CP violation through $B$-meson
decays is the direct determination of the angle $\gamma$ of
the unitarity triangle in a variety of ways. We give a brief
overview of the implications of the current $B$-factory data 
for flavour-symmetry strategies, and discuss new, theoretically
clean methods employing pure tree decays of neutral $B_{d,s}$ mesons.
\end{quotation}

\vspace{2.5cm}

\begin{center} 
{\sl Invited talk at the Workshop on the CKM Unitarity Triangle,\\ 
Durham, United Kingdom, 5--9 April 2003\\
To appear in the Proceedings (SLAC eConf service)}
\end{center}

\vfill
\noindent
CERN-TH/2003-141\\
June 2003

\end{titlepage}

\thispagestyle{empty}
\vbox{}
\newpage
 
\setcounter{page}{1}


\begin{abstract}One of the key elements in the testing of the 
Standard-Model description of CP violation through $B$-meson
decays is the direct determination of the angle $\gamma$ of
the unitarity triangle in a variety of ways. We give a brief
overview of the implications of the current $B$-factory data 
for flavour-symmetry strategies, and discuss new, theoretically
clean methods employing pure tree decays of neutral $B_{d,s}$ mesons.
\end{abstract}

\maketitle


%
%
%
\section{Introduction}\label{sec:intro}
As is well known, the ``standard analysis'' of the unitarity triangle 
(UT) allows us to obtain ``indirect'' ranges for its three angles $\alpha$, 
$\beta$ and $\gamma$ \cite{CKMyellow2002}. To this end, we apply the Standard 
Model (SM), and measure the UT side $R_b\propto |V_{ub}/V_{cb}|$ through 
semileptonic $B$ decays caused by $b\to u\ell\overline{\nu}_{\ell}, 
c\ell\overline{\nu}_{\ell}$ quark-level transitions, determine the UT side 
$R_t\propto |V_{td}/V_{cb}|$ through experimental information on the mass 
differences $\Delta M_{d,s}$, and employ the indirect CP violation in the 
neutral kaon system, which is described by the famous observable 
$\varepsilon_K$, to calculate a hyperbola in the 
$\overline{\rho}$--$\overline{\eta}$ plane of the
generalized Wolfenstein parameters \cite{wolf,BLO}. 

A crucial element in the stringent testing of the Kobayashi--Maskawa
mechanism for CP violation is the direct determination of the UT angle
$\gamma$. Many strategies to accomplish this task were proposed, where
also decays of $B_s$ mesons -- the ``El Dorado'' for hadron colliders -- 
play a prominent r\^ole (for a detailed review, see \cite{RF-PHYS-REP}). 
The main goal of this programme is to overconstrain $\gamma$ as much as 
possible, hoping to encounter inconsistencies.

\section{Flavour-Symmetry Strategies}\label{sec:flavour}
A very interesting avenue to determine $\gamma$ from non-leptonic
$B$ decays is provided by the flavour symmetries of strong interactions.
The basic idea is to use isospin or $SU(3)$ arguments to deal with
``unknown'' hadronic matrix elements of local four-quark operators. 
In some cases, also plausible dynamical assumptions have to be made
in order to reduce the number of parameters entering the decay 
amplitudes.

\subsection{$B\to\pi K$ Decays}
These modes originate from $\overline{b}\to\overline{d}d\overline{s},
\overline{u}u\overline{s}$ quark-level processes, and may receive 
contributions both from penguin and from tree topologies, where the latter 
bring $\gamma$ into the game. Since the ratio of tree to penguin
contributions is governed by the tiny CKM factor 
$|V_{us}V_{ub}^\ast/(V_{ts}V_{tb}^\ast)|\approx0.02$, $B\to\pi K$ decays 
are dominated by QCD penguins, despite their loop suppression. 
As far as electroweak (EW) penguins are concerned, their effects are 
expected to be negligible in the case of the $B^0_d\to\pi^-K^+$, 
$B^+\to\pi^+K^0$ system, as they contribute here only in colour-suppressed 
form. On the other hand, EW penguins may also contribute in colour-allowed 
form to $B^+\to\pi^0K^+$ and $B^0_d\to\pi^0K^0$, and are hence expected 
to be sizeable in these modes, i.e.\ of the same order of magnitude as 
the tree topologies.

Thanks to interference effects between tree and penguin amplitudes, 
we obtain sensitivity on $\gamma$. In order to determine this angle, 
we may use an isospin relation as a starting point, suggesting the 
following combinations: the ``mixed'' $B^\pm\to\pi^\pm K$, 
$B_d\to\pi^\mp K^\pm$ system \cite{PAPIII}--\cite{defan}, the ``charged'' 
$B^\pm\to\pi^\pm K$, $B^\pm\to\pi^0K^\pm$ system 
\cite{NR}--\cite{BF-neutral1}, and the ``neutral'' 
$B_d\to\pi^0 K$, $B_d\to\pi^\mp K^\pm$ system \cite{BF-neutral1,BF-neutral2}.
As noted in \cite{BF-neutral1}, all three $B\to\pi K$ systems can be described 
by the same set of formulae by just making straightforward replacements of 
variables. Let us first focus on the charged and neutral $B\to\pi K$ systems. 
In order to determine $\gamma$ and the corresponding strong phases, we have 
to introduce appropriate CP-conserving and CP-violating observables:
\begin{equation}\label{charged-obs}
\left.\begin{array}{l}R_{\rm c}\\A_0^{\rm c}\end{array}\right.
\equiv2\left[\frac{\mbox{BR}(B^+\to\pi^0K^+)\pm
\mbox{BR}(B^-\to\pi^0K^-)}{\mbox{BR}(B^+\to\pi^+K^0)+
\mbox{BR}(B^-\to\pi^-\overline{K^0})}\right]
\end{equation}
\begin{equation}\label{neutral-obs}
\left.\begin{array}{l}R_{\rm n}\\A_0^{\rm n}\end{array}\right.
\equiv\frac{1}{2}\left[\frac{\mbox{BR}(B^0_d\to\pi^-K^+)\pm
\mbox{BR}(\overline{B^0_d}\to\pi^+K^-)}{\mbox{BR}(B^0_d\to\pi^0K^0)+
\mbox{BR}(\overline{B^0_d}\to\pi^0\overline{K^0})}\right],
\end{equation}
where the $R_{\rm c,n}$ and $A_0^{\rm c,n}$ refer to the plus and 
minus signs, respectively. For the parametrization of these observables, 
we employ the isospin relation mentioned above, and assume that certain 
rescattering effects are small, which is in accordance with the 
QCD factorization picture \cite{BBNS1,BBNS3}; large rescattering processes 
would be indicated by $B\to KK$ modes, which are already strongly constrained 
by the $B$ factories, and could be included through more elaborate 
strategies \cite{defan,neubert-BpiK,BF-neutral1}. Following these lines, 
we may write
\begin{equation}
R_{\rm c,n}=\mbox{fct}(q,r_{\rm c,n},\delta_{\rm c,n},\gamma), \quad
A_0^{\rm c,n}=\mbox{fct}(r_{\rm c,n},\delta_{\rm c,n},\gamma),
\end{equation}
where the parameters $q$, $r_{\rm c,n}$ and $\delta_{\rm c,n}$ have 
the following meaning: $q$ describes the ratio of EW penguin to tree 
contributions, and can be determined with the help of $SU(3)$ 
flavour-symmetry arguments, yielding $q\sim 0.7$ \cite{NR}. On the
other hand, $r_{\rm c,n}$ measures the ratio of tree to QCD penguin 
topologies, and can be fixed through $SU(3)$ arguments and data on 
$B^\pm\to\pi^\pm\pi^0$ modes \cite{GRL}, which give $r_{\rm c,n}\sim0.2$. 
Finally, $\delta_{\rm c,n}$ is the CP-conserving strong phase between the 
tree and QCD penguin amplitudes. Since we may fix $q$ and $r_{\rm c,n}$, 
the observables $R_{\rm c,n}$ and $A_0^{\rm c,n}$ actually depend only 
on the two ``unknown'' parameters $\delta_{\rm c,n}$ and $\gamma$. If we 
vary them within their allowed ranges, i.e.\ 
$-180^\circ\leq \delta_{\rm c,n}\leq+180^\circ$ and 
$0^\circ\leq \gamma \leq180^\circ$, we obtain an allowed region in the 
$R_{\rm c,n}$--$A_0^{\rm c,n}$ plane \cite{FlMa1,FlMa2}. Should the
measured values of $R_{\rm c,n}$ and $A_0^{\rm c,n}$ fall outside this 
region, we would have an immediate signal for new physics (NP). On the 
other hand, should the measurements lie inside the allowed range, 
$\gamma$ and $\delta_{\rm c,n}$ could be extracted. The value of $\gamma$ 
thus obtained could then be compared with the results of other
strategies, whereas the strong phase $\delta_{\rm c,n}$ would offer
interesting insights into hadron dynamics. This exercise can be performed 
separately for the charged and neutral $B\to\pi K$ systems.

\begin{figure}
\vspace*{-0.0cm}
$$\hspace*{-0.3cm}
\epsfysize=0.15\textheight
\epsfxsize=0.17\textheight
\epsffile{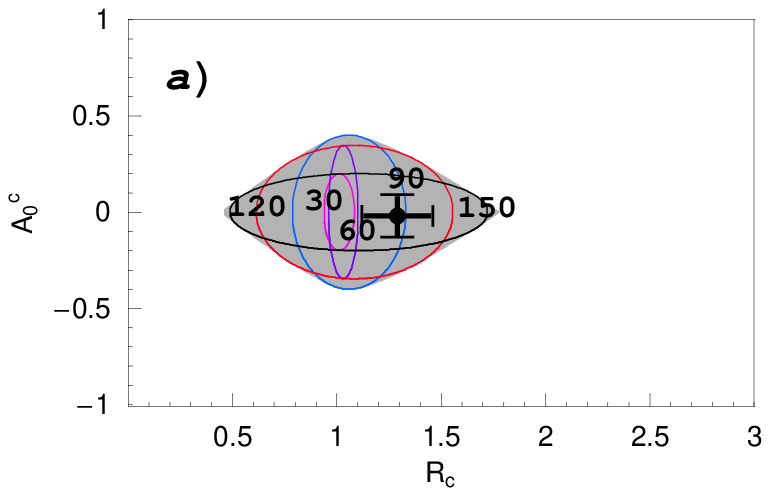} \hspace*{0.3cm}
\epsfysize=0.15\textheight
\epsfxsize=0.17\textheight
\epsffile{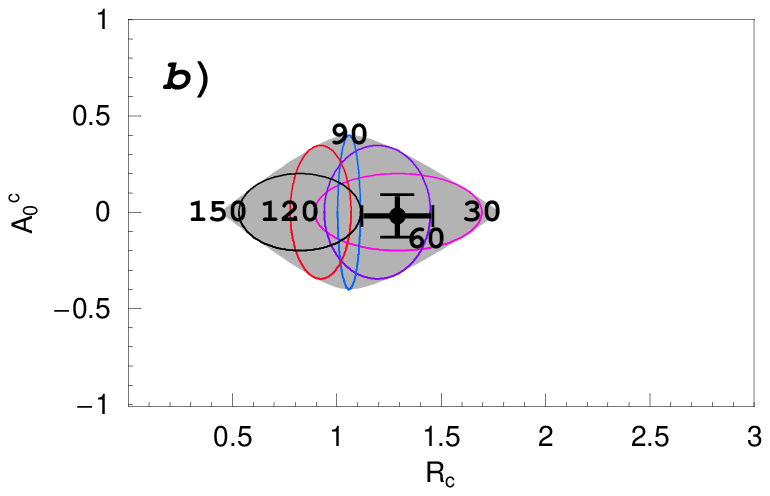}
$$
\vspace*{-0.6cm}
$$\hspace*{-0.3cm}
\epsfysize=0.15\textheight
\epsfxsize=0.17\textheight
\epsffile{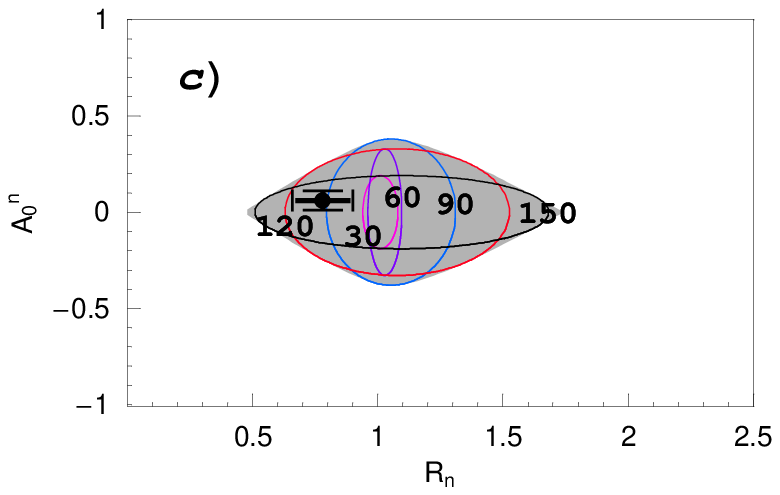} \hspace*{0.3cm}
\epsfysize=0.15\textheight
\epsfxsize=0.17\textheight
\epsffile{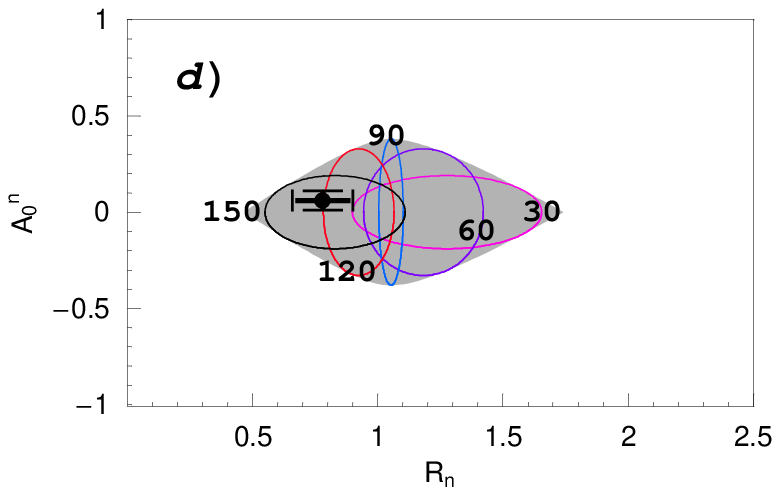}
$$
\vspace*{-0.8truecm}
\caption{The allowed regions in observable space of the charged
($r_{\rm c}=0.20$; (a), (b)) and neutral ($r_{\rm n}=0.19$; (c), (d))
$B\to \pi K$ systems for $q=0.68$: in (a) and (c), we show also the 
contours for fixed values of $\gamma$, whereas we give the curves 
arising for fixed values of $|\delta_{\rm c}|$ and $|\delta_{\rm n}|$ 
in (b) and (d), respectively.}\label{fig:BpiK-OS}
\end{figure}

In Fig.~\ref{fig:BpiK-OS}, we show the allowed regions in the 
$R_{\rm c,n}$--$A_0^{\rm c,n}$ planes \cite{FlMa2}, where the 
crosses represent the averages of the most recent $B$-factory 
data \cite{CLEO-III,olsen}. As can be read off from the contours 
in these figures, both the charged and the
neutral $B\to \pi K$ data favour $\gamma\gsim90^\circ$, which 
would be in conflict with the CKM fits resulting from the ``standard
analysis'' of the UT, favouring
\begin{equation}\label{SM-gam}
50^\circ\leq\gamma\leq70^\circ.
\end{equation}
Interestingly, the charged modes point towards $|\delta_{\rm c}|\lsim90^\circ$ 
(factorization predicts $\delta_{\rm c}$ to be close to 
$0^\circ$ \cite{BBNS3}), whereas the neutral decays seem to prefer 
$|\delta_{\rm n}|\gsim90^\circ$. Since we do not expect $\delta_{\rm c}$ 
to differ significantly from $\delta_{\rm n}$, we
arrive at a ``puzzling'' picture of the kind that was already considered
a couple of years ago in \cite{BF-neutral2}. Unfortunately, the 
experimental uncertainties do not yet allow us to draw definite 
conclusions. As far as the mixed $B\to\pi K$ system is concerned, the 
data fall well into the SM region in observable space, and do not show
any ``anomalous'' behaviour at the moment. For a selection of alternative 
analyses of $B\to\pi K$ decays, see 
\cite{BBNS3,PQCD-appl,matias,ital-corr,GR-BpiK-recent}.

\subsection{The $B_d\to\pi^+\pi^-$, $B_s\to K^+K^-$ System}
The decay $B_d^0\to\pi^+\pi^-$ originates from $\overline{b}\to
\overline{u}u\overline{d}$ quark-level processes. Within the SM, 
its transition amplitude may be written as
\begin{equation}\label{Bpipi-ampl}
A(B_d^0\to\pi^+\pi^-)\propto\left[e^{i\gamma}-de^{i\theta}\right],
\end{equation}
where the CP-conserving strong parameter $d e^{i\theta}$ measures --
sloppily speaking -- the 
ratio of the penguin to tree contributions \cite{RF-BsKK}. As is
well known, if we had negligible penguin contributions, i.e.\ 
$d=0$, the direct and mixing-induced CP asymmetries provided by the
time-dependent rate asymmetry of the kind \cite{RF-PHYS-REP}
\begin{eqnarray}
\lefteqn{\frac{\Gamma(B^0_q(t)\to f)-
\Gamma(\overline{B^0_q}(t)\to f)}{\Gamma(B^0_q(t)\to f)+
\Gamma(\overline{B^0_q}(t)\to f)}}\nonumber\\
&&=\left[\frac{{\cal A}_{\rm CP}^{\rm dir}\,\cos(\Delta M_q t)+
{\cal A}_{\rm CP}^{\rm mix}\,\sin(\Delta 
M_q t)}{\cosh(\Delta\Gamma_qt/2)-{\cal A}_{\rm 
\Delta\Gamma}\,\sinh(\Delta\Gamma_qt/2)}\right]\label{ACP-time}
\end{eqnarray}
were simply given as follows:
\begin{equation}\label{Adir-d0}
{\cal A}_{\rm CP}^{\rm dir}(B_d\to\pi^+\pi^-)=0
\end{equation}
\begin{equation}\label{Amix-d0}
{\cal A}_{\rm CP}^{\rm mix}(B_d\to\pi^+\pi^-)=\sin(\phi_d+2\gamma)
\stackrel{\rm SM}{=}-\sin 2\alpha.
\end{equation}
Here we have used both the SM expression $\phi_d=2\beta$ and the 
unitarity relation $2\beta+2\gamma=2\pi-2\alpha$ to derive the last 
identity in (\ref{Amix-d0}). We observe that $\phi_d$ and $\gamma$ actually 
enter directly ${\cal A}_{\rm CP}^{\rm mix}(B_d\to\pi^+\pi^-)$, and not 
$\alpha$. Consequently, since we may fix $\phi_d$ straightforwardly 
through $B_d\to J/\psi K_{\rm S}$, also if NP should contribute to
$B^0_d$--$\overline{B^0_d}$ mixing, we may use the CP-violating
$B_d\to\pi^+\pi^-$ observables to probe $\gamma$. This avenue is of 
great advantage to deal with penguin effects and possible NP contributions 
to $B^0_d$--$\overline{B^0_d}$ mixing \cite{FlMa2,RF-BsKK,RF-Bpipi}.

Measurements of the CP asymmetries of the $B_d\to\pi^+\pi^-$ channel
are already available:
\begin{equation}\label{Adir-exp}
{\cal A}_{\rm CP}^{\rm dir}=\left\{
\begin{array}{ll}
-0.30\pm0.25\pm0.04 & \mbox{(BaBar \cite{BaBar-Bpipi})}\\
-0.77\pm0.27\pm0.08 & \mbox{(Belle \cite{Belle-Bpipi})}
\end{array}
\right.
\end{equation}
\begin{equation}\label{Amix-exp}
{\cal A}_{\rm CP}^{\rm mix}=\left\{
\begin{array}{ll}
-0.02\pm0.34\pm0.05& \mbox{(BaBar \cite{BaBar-Bpipi})}\\
+1.23\pm0.41 ^{+0.07}_{-0.08} & \mbox{(Belle \cite{Belle-Bpipi}).}
\end{array}
\right.
\end{equation}
The BaBar and Belle results are unfortunately not fully consistent with
each other. If we form, nevertheless, the weighted averages of 
(\ref{Adir-exp}) and (\ref{Amix-exp}), applying the rules of the Particle 
Data Group (PDG), we obtain 
\begin{eqnarray}
{\cal A}_{\rm CP}^{\rm dir}
&=&-0.51\pm0.19 \,\, (0.23)
\label{Bpipi-CP-averages}\\
{\cal A}_{\rm CP}^{\rm mix}
&=&+0.49\pm0.27 \,\, (0.61),
\label{Bpipi-CP-averages2}
\end{eqnarray}
where the errors in brackets are those increased by the PDG scaling-factor 
procedure \cite{PDG}. Direct CP violation at this level would require large 
penguin contributions, with large CP-conserving strong phases, which are
not suggested by the QCD factorization approach, pointing towards
${\cal A}_{\rm CP}^{\rm dir}(B_d\to\pi^+\pi^-)\sim +0.1$ \cite{BBNS3}.
In addition to (\ref{Bpipi-CP-averages}), a significant impact of penguins 
on $B_d\to\pi^+\pi^-$ is also indicated by the data for the 
$B\to\pi K,\pi\pi$ branching ratios \cite{FlMa2,RF-Bpipi}, as well as by 
theoretical considerations \cite{BBNS3,PQCD-appl}. Consequently, it is 
already evident that we {\it must} care about the penguin contributions  
in order to extract information on the UT from the $B_d\to\pi^+\pi^-$
CP asymmetries. 

\begin{figure}
\centerline{
\rotate[r]{
\epsfxsize=5.5truecm
{\epsffile{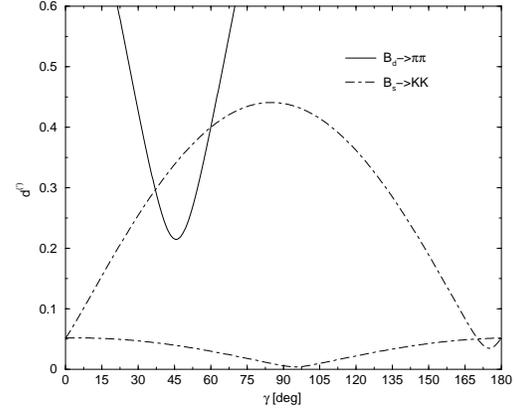}}}}
\caption{The $\gamma$--$d^{(')}$ contours for an example with
$d=d'=0.4$, $\theta=\theta'=140^\circ$, $\phi_d=47^\circ$, 
$\phi_s=0^\circ$, $\gamma=60^\circ$, corresponding to
${\cal A}_{\rm CP}^{\rm dir}(B_d\to\pi^+\pi^-)=-0.30$,
${\cal A}_{\rm CP}^{\rm mix}(B_d\to\pi^+\pi^-)=+0.63$,
${\cal A}_{\rm CP}^{\rm dir}(B_s\to K^+K^-)=+0.16$ and
${\cal A}_{\rm CP}^{\rm mix}(B_s\to K^+K^-)=-0.17$.}\label{fig:gam-d}
\end{figure}

In the literature, many approaches to address this challenging problem 
can be found (see \cite{RF-PHYS-REP} and references therein). Let us
focus here on a strategy that complements $B_d\to\pi^+\pi^-$
with $B_s\to K^+K^-$ \cite{RF-BsKK}. In analogy to (\ref{Bpipi-ampl}), the 
amplitude of the latter decay, which is very accessible at $B$-decay 
experiments at hadron colliders \cite{TEV-BOOK}--\cite{vagnoni}, can 
be written as follows:
\begin{equation}\label{BsKK-ampl}
A(B_s^0\to K^+K^-)\propto\left[e^{i\gamma}+\left(
\frac{1-\lambda^2}{\lambda^2}\right)d'e^{i\theta'}\right],
\end{equation}
where the hadronic quantities $d'$ and $\theta'$ are the $B_s\to K^+K^-$
counterparts of the parameters $d$ and $\theta$. Consequently, we
obtain observables of the following structure:
\begin{equation}\label{Bpipi-obs}
\begin{array}{rcl}
{\cal A}_{\rm CP}^{\rm dir}(B_d\to\pi^+\pi^-)&=&
\mbox{fct}(d,\theta,\gamma)\\
{\cal A}_{\rm CP}^{\rm mix}(B_d\to\pi^+\pi^-)&=&
\mbox{fct}(d,\theta,\gamma,\phi_d)
\end{array}
\end{equation}
\begin{equation}\label{BsKK-obs}
\begin{array}{rcl}
{\cal A}_{\rm CP}^{\rm dir}(B_s\to K^+K^-)&=&
\mbox{fct}(d',\theta',\gamma)\\
{\cal A}_{\rm CP}^{\rm mix}(B_s\to K^+K^-)&=&
\mbox{fct}(d',\theta',\gamma,\phi_s),
\end{array}
\end{equation}
where $\phi_d$ and $\phi_s$ entering the mixing-induced CP asymmetries
can straightforwardly be fixed separately \cite{RF-PHYS-REP}, also if NP 
should contribute to $B^0_q$--$\overline{B^0_q}$ mixing ($q\in\{d,s\}$).
We may then use the $B_d\to\pi^+\pi^-$ observables to eliminate the
strong phase $\theta$, which allows us to determine $d$ as a 
function of $\gamma$ in a {\it theoretically clean} manner, i.e.\ 
without using flavour symmetry or plausible dynamical assumptions. 
Analogously, we may employ the $B_s\to K^+K^-$ CP asymmetries to 
eliminate $\theta'$, yielding $d'$ as a {\it theoretically clean} 
function of $\gamma$. If we have a look at the corresponding Feynman 
diagrams, we observe that $B_d\to\pi^+\pi^-$ is related to $B_s\to K^+K^-$ 
through an interchange of all down and strange quarks. Consequently, 
the $U$-spin flavour symmetry of strong interactions implies
\begin{equation}\label{U-spin-rel}
d'=d, \quad \theta'=\theta. 
\end{equation}
If we now apply the former relation, we may determine $\gamma$ and $d$ from 
the theoretically clean contours in the $\gamma$--$d$ and $\gamma$--$d'$ 
planes, as well as the strong phases $\theta'$ and $\theta$, which provide 
a nice consistency check of the latter $U$-spin relation \cite{RF-BsKK}. 
In Fig.~\ref{fig:gam-d}, we have illustrated these contours for a 
specific example.

This strategy is very promising from an experimental point of view: 
at run II of the Tevatron and the LHC, experimental accuracies for 
$\gamma$ of ${\cal O}(10^\circ)$ and ${\cal O}(1^\circ)$, 
respectively, are expected \cite{TEV-BOOK,LHC-BOOK}. As far as 
$U$-spin-breaking corrections to $d'=d$ are concerned \cite{RF-BsKK,beneke}, 
they enter the determination of $\gamma$ through a relative shift of
the $\gamma$--$d$ and $\gamma$--$d'$ contours; their impact on the 
extracted value of $\gamma$ depends on the form of these curves, which 
is fixed through the measured observables. In the examples discussed in 
\cite{RF-PHYS-REP,RF-BsKK}, as well as in the one shown in 
Fig.~\ref{fig:gam-d}, the extracted value of $\gamma$ would be very 
robust under such corrections.

\begin{figure}[t]
$$\hspace*{-1.cm}
\epsfysize=0.2\textheight
\epsfxsize=0.3\textheight
\epsffile{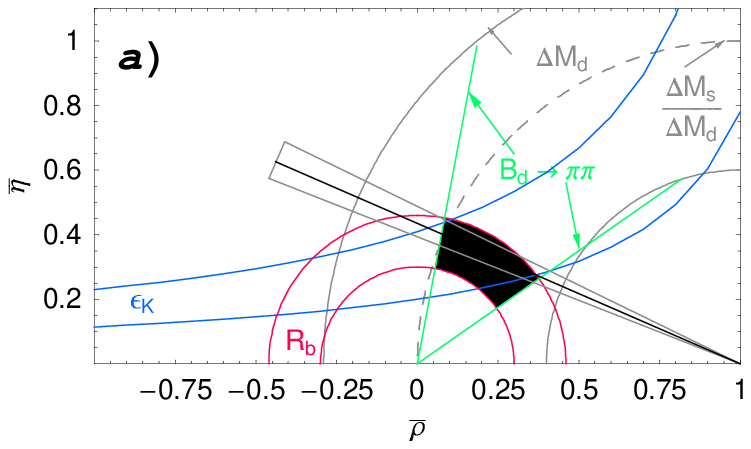} 
$$
\vspace*{0.0cm}
$$\hspace*{-1.cm}
\epsfysize=0.2\textheight
\epsfxsize=0.3\textheight
\epsffile{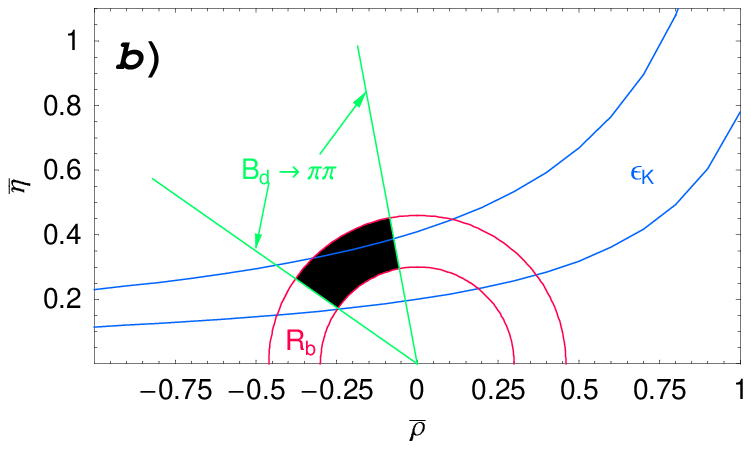}
$$
\caption{The allowed regions for the UT fixed through $R_b$ and CP 
violation in $B_d\to\pi^+\pi^-$, as described in the text: (a) and (b) 
correspond to $\phi_d=47^\circ$ and $\phi_d=133^\circ$, respectively 
($H=7.5$).}\label{fig:rho-eta-Bpipi}
\end{figure}

Since $B_s\to K^+K^-$ is not accessible at the $\Upsilon(4S)$ $e^+e^-$ 
$B$ factories, we may not yet implement this strategy. However, 
$B_s\to K^+K^-$ is related to $B_d\to\pi^\mp K^\pm$ through an interchange 
of spectator quarks. Consequently, we may approximately replace 
$B_s\to K^+K^-$ through $B_d\to\pi^\mp K^\pm$ to deal with the penguin 
problem in $B_d\to\pi^+\pi^-$ \cite{RF-Bpipi}. The utility of 
$B_d\to\pi^\mp K^\pm$ decays to control the penguin effects in
$B_d\to\pi^+\pi^-$ was also emphasized in \cite{siwo1}. In order to 
explore the implications of the $B$-factory data, the following quantity, 
which involves the ratio of the corresponding CP-averaged branching ratios, 
plays a key r\^ole:
\begin{equation}\label{H-det}
H=\frac{1}{\epsilon}\left(\frac{f_K}{f_\pi}\right)^2
\left[\frac{\mbox{BR}(B_d\to\pi^+\pi^-)}{\mbox{BR}(B_d\to\pi^\mp K^\pm)}
\right],
\end{equation}
where $\epsilon\equiv\lambda^2/(1-\lambda^2)$. The current experimental
status of $H$ is given by
\begin{equation}\label{H-exp}
H=\left\{\begin{array}{ll}
7.4\pm2.5 & \mbox{(CLEO \cite{CLEO-III})}\\
7.8\pm1.2 & \mbox{(BaBar \cite{olsen})}\\
7.1\pm1.2 & \mbox{(Belle \cite{olsen}).}
\end{array}\right.
\end{equation}
If we apply (\ref{U-spin-rel}), we may write
\begin{equation}\label{H-expr}
H=\mbox{fct}(d,\theta,\gamma).
\end{equation}
Consequently, (\ref{Bpipi-obs}) and (\ref{H-expr}) provide sufficient 
information to determine $\gamma$, $d$ and $\theta$ \cite{RF-BsKK,RF-Bpipi}. 
As discussed in detail in \cite{FlMa2}, if we follow these lines and 
complement the twofold solution $\phi_d\sim47^\circ\lor 133^\circ$, which
is implied by the measured mixing-induced CP violation in 
$B_d\to J/\psi K_{\rm S}$, with (\ref{Bpipi-CP-averages}),
(\ref{Bpipi-CP-averages2}) and (\ref{H-exp}), we obtain the following
picture: in the case of $\phi_d=47^\circ$, the data point towards 
$\gamma\sim60^\circ$, so that not only $\phi_d$ would be in 
accordance with the CKM fits, but also $\gamma$. On the other hand, for 
$\phi_d=133^\circ$, the experimental values favour $\gamma\sim120^\circ$. 
At first sight, this may look puzzling. However, since the $\phi_d=133^\circ$ 
solution would definitely require NP contributions to 
$B^0_d$--$\overline{B^0_d}$ mixing, we may no longer use the SM 
interpretation of $\Delta M_d$ in this case to fix the UT side 
$R_t$, which is a crucial ingredient for the $\gamma$ range in 
(\ref{SM-gam}). Consequently, if we choose $\phi_d=133^\circ$, $\gamma$ 
may well be larger than $90^\circ$. As we have alread noted, the 
$B\to\pi K$ data seem to favour such values; a similar feature is also 
suggested by the observed small $B_d\to\pi^+\pi^-$ rate \cite{FlMa2,RF-Bpipi}. 
Interestingly, the measured branching ratio for the rare kaon decay 
$K^+\to\pi^+\nu\overline{\nu}$ seems to point towards 
$\gamma>90^\circ$ as well \cite{dAI,isidori}, thereby also favouring the 
unconventional solution of $\phi_d=133^\circ$ \cite{FIM}. Further 
valuable information on this exciting possibility can be obtained in
the future from the rare decays $B_{s,d}\to\mu^+\mu^-$. 

We could straightforwardly accommodate this picture in a NP scenario, 
where we have large effects in $B^0_d$--$\overline{B^0_d}$ mixing, 
but not in the $\Delta B=1$ and $\Delta S=1$ decay processes. 
Such kind of NP was already considered several years ago \cite{GNW,siwo2}, 
and can be motivated by generic arguments and within supersymmetry \cite{FIM}.
Since the determination of $R_b$ through semileptonic tree decays is in 
general very robust under NP effects and would not be affected either in 
this particular scenario, we may complement $R_b$ with the range for 
$\gamma$ extracted from our $B_d\to\pi^+\pi^-$ analysis, allowing us to 
fix the apex of the UT in the $\overline{\rho}$--$\overline{\eta}$ plane. 
The results of this exercise are summarized in Fig.~\ref{fig:rho-eta-Bpipi}, 
following \cite{FIM}, where also numerical values for $\alpha$, $\beta$ and 
$\gamma$ are given and a detailed discussion of the theoretical uncertainties 
can be found (see also \cite{matias-proc}). Note that the SM contours 
implied by $\Delta M_d$, which are included in Fig.~\ref{fig:rho-eta-Bpipi} 
(a) to guide the eye, are absent in (b) since, there, 
$B^0_d$--$\overline{B^0_d}$ mixing would receive NP contributions. 
In this case also, we may no 
longer simply represent $\phi_d$ by a straight line, as the one in 
Fig.~\ref{fig:rho-eta-Bpipi} (a), which corresponds to 
$\phi_d\stackrel{{\rm SM}}{=}2\beta$, since we would now have 
$\phi_d=2\beta+\phi_d^{\rm NP}$, with $\phi_d^{\rm NP}\not=0^\circ$. 
However, we may easily read off the ``correct'' value of $\beta$ from the 
black region in Fig.~\ref{fig:rho-eta-Bpipi} (b) \cite{FIM}. Interestingly, 
both black regions in Fig.~\ref{fig:rho-eta-Bpipi} (a) and (b) are consistent 
with the SM $\varepsilon_K$ hyperbola.

\begin{figure}
\centerline{{
\epsfysize=4.6truecm
{\epsffile{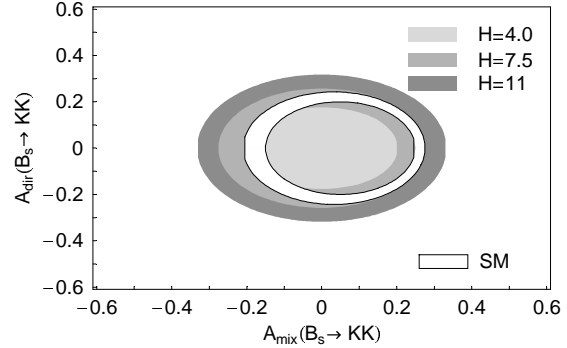}}}}
\caption{The allowed regions in the $B_s\to K^+K^-$ observable space 
originating for $\phi_s=0^\circ$ and for various values of $H$. The SM 
region appears if we restrict $\gamma$ to 
(\ref{SM-gam}) and choose $H=7.5$.}\label{fig:BsKK-obs-space}
\end{figure}

Since the current experimental status of the CP-violating $B_d\to\pi^+\pi^-$ 
observables is unsatisfactory, we may not yet draw definite conclusions from 
this analysis, although it illustrates nicely how the corresponding strategy 
works. However, the $B$-factory measurements of CP violation in 
$B_d\to\pi^+\pi^-$ will improve significantly in the future, thereby 
providing more stringent constraints on $\gamma$, which we may then
convert into narrower ranges for the apex of the UT. Another milestone in 
this programme is the measurement of the CP-averaged $B_s\to K^+K^-$ 
branching ratio at run II of the Tevatron, which will allow a much better 
determination of $H$ that no longer relies on dynamical assumptions. 
Finally, if also the direct and mixing-induced CP asymmetries of 
$B_s\to K^+K^-$ are measured, we may determine $\gamma$ through a minimal 
$U$-spin input, as we have sketched above. Interestingly, $H$ implies 
already a very narrow SM target region in the space of the CP-violating 
$B_s\to K^+K^-$ observables, as can be seen in Fig.~\ref{fig:BsKK-obs-space}. 
It will be exciting to see whether the data will actually fall into the 
small elliptical region in this figure. After important steps by the CDF 
collaboration, LHCb and BTeV should be able to fully exploit the rich 
physics potential of the $B_d\to\pi^+\pi^-$, $B_s\to K^+K^-$ system. 

There are several other promising $B_s$ decays, which we shall address in 
the discussion of the following section.

\section{Theoretically Clean Strategies}\label{sec:clean}
In order to determine $\gamma$ in a theoretically clean manner,
pure tree decays of $B$ mesons play the central r\^ole. There are 
basically two kinds of approaches: 
\begin{itemize}
\item[i] We may induce interference effects in $B$ decays through
subsequent decays of neutral $D$ mesons, satisfying 
$D^0,\overline{D^0}\to f_D$. Important examples are 
$B^\pm\to D K^\pm$, $B_c^\pm\to D D_s^\pm$, ...\ modes 
\cite{gw}--\cite{MG}.
\item[ii] We may employ neutral $B_q$ modes, where both
$B^0_q$ and $\overline{B^0_q}$ may decay into the same final state,
yielding interference between mixing and decay processes, e.g.\
$B_d\to D^{(\ast)\pm}\pi^\mp$, $B_s\to D_s^{(\ast)\pm} K^\mp$, ...\
\cite{DuSa}--\cite{BsDsK}. 
\end{itemize}
In this section, we will focus on two kinds of new strategies, which
were proposed in \cite{RF-gam-eff-03,RF-gam-det-03} and \cite{RF-gam-ca}.

\subsection{$B_d\to D K_{\rm S(L)}$, $B_s\to D\eta^{(')}, 
D\phi$, ...}\label{ssec:BqDfr}
Colour-suppressed $B^0_d\to D^0 K_{\rm S}$ decays and similar modes 
provide interesting tools to explore CP violation \cite{GroLo,KayLo}. 
In the following, we consider $B^0_q\to D^0f_r$ transitions, 
where $r\in\{s,d\}$ distinguishes between $b\to Ds$ and $b\to D d$ 
processes \cite{RF-gam-eff-03,RF-gam-det-03}. If we require 
$({\cal CP})|f_r\rangle=\eta_{\rm CP}^{f_r}|f_r\rangle$,
$B^0_q$ and $\overline{B^0_q}$ mesons may both decay into $D^0f_r$, 
thereby leading to interference effects between $B^0_q$--$\overline{B^0_q}$ 
mixing and decay processes, which involve the weak phase $\phi_q+\gamma$
(see ii): 
\begin{itemize}
\item For $r=s$, i.e.\ $B_d\to D K_{\rm S(L)}$, 
$B_s\to D\eta^{(')}, D\phi$, ...\ modes, these interference effects are 
governed by a hadronic parameter 
$x_{f_s}e^{i\delta_{f_s}}\propto R_b\approx0.4$, and are hence 
favourably large. 
\item For $r=d$, i.e.\ $B_s\to DK_{\rm S(L)}$, $B_d\to D\pi^0, D\rho^0$ ...\ 
modes, the interference effects are tiny because of 
$x_{f_d}e^{i\delta_{f_d}}\propto -\lambda^2R_b\approx -0.02$. 
\end{itemize}

Let us first focus on the $r=s$ case. If we consider $B_q\to D_\pm f_s$ 
modes, where $({\cal CP})|D_\pm\rangle=\pm|D_\pm\rangle$, additional 
interference effects between $B_q^0\to D^0 f_s$ and 
$B_q^0\to \overline{D^0} f_s$ arise at the decay level, involving $\gamma$
(see i). The most straightforward observable is the ``untagged'' rate
\begin{eqnarray}
\lefteqn{\langle\Gamma(B_q(t)\to D_\pm f_s)\rangle\equiv
\Gamma(B^0_q(t)\to D_\pm f_s)+
\Gamma(\overline{B^0_q}(t)\to D_\pm f_s)}\nonumber\\
&&\stackrel{\Delta\Gamma_q=0}{=}
\left[\Gamma(B^0_q\to D_\pm f_s)+
\Gamma(\overline{B^0_q}\to D_\pm f_s)\right]e^{-\Gamma_qt}\nonumber\\
&&\mbox{}~~~\equiv \langle\Gamma(B_q\to D_\pm f_s)\rangle e^{-\Gamma_qt},
\end{eqnarray}
providing the following ``untagged'' rate asymmetry:
\begin{equation}
\Gamma_{+-}^{f_s}\equiv
\frac{\langle\Gamma(B_q\to D_+ f_s)\rangle-\langle
\Gamma(B_q\to D_- f_s)\rangle}{\langle\Gamma(B_q\to D_+ f_s)\rangle
+\langle\Gamma(B_q\to D_- f_s)\rangle}.
\end{equation}
Interestingly, already the quantity $\Gamma_{+-}^{f_s}$ offers valuable 
information about $\gamma$ \cite{RF-gam-eff-03}, since bounds on this 
angle are implied by
\begin{equation}
|\cos\gamma|\geq |\Gamma_{+-}^{f_s}|. 
\end{equation}
Moreover, if we take into account that the factorization picture suggests 
$\cos\delta_{f_s}>0$ \cite{RF-gam-det-03}, we obtain 
\begin{equation}\label{sgn-cos-gam}
\mbox{sgn}(\cos\gamma)=\mbox{sgn}(\Gamma_{+-}^{f_s}),
\end{equation} 
i.e.\ we may decide whether $\gamma$ is smaller or larger than 
$90^\circ$.
If we employ, in addition, the mixing-induced observables 
$S_\pm^{f_s}\equiv {\cal A}_{\rm CP}^{\rm mix}(B_q\to D_\pm f_s)$, 
we may determine $\gamma$. To this end, it is convenient to introduce
the quantities
\begin{equation}
\langle S_{f_s}\rangle_\pm\equiv\left(S_+^{f_s}\pm S_-^{f_s}\right)/2.
\end{equation}
Expressing the $\langle S_{f_s}\rangle_\pm$ in terms of the $B_q\to D_\pm f_s$ 
decay parameters gives rather complicated formulae. However, 
complementing the $\langle S_{f_s}\rangle_\pm$ with $\Gamma_{+-}^{f_s}$ 
yields 
\begin{equation}\label{key-rel}
\tan\gamma\cos\phi_q=
\left[\frac{\eta_{f_s} \langle S_{f_s}
\rangle_+}{\Gamma_{+-}^{f_s}}\right]+\left[\eta_{f_s}\langle S_{f_s}\rangle_--
\sin\phi_q\right],
\end{equation}
where $\eta_{f_s}\equiv(-1)^L\eta_{\rm CP}^{f_s}$, with $L$ denoting the 
$Df_s$ angular momentum \cite{RF-gam-eff-03}. Using this simple -- 
but {\it exact} -- relation, we obtain the twofold solution 
$\gamma=\gamma_1\lor\gamma_2$, with $\gamma_1\in[0^\circ,180^\circ]$ and 
$\gamma_2=\gamma_1+180^\circ$. Since $\cos\gamma_1$ and $\cos\gamma_2$
have opposite signs, (\ref{sgn-cos-gam}) allows us to fix $\gamma$ 
{\it unambiguously}. Another advantage of (\ref{key-rel}) is that 
$\langle S_{f_s}\rangle_+$ and $\Gamma_{+-}^{f_s}$ are both proportional 
to $x_{f_s}\approx0.4$, so that the first term in square brackets is 
of ${\cal O}(1)$, whereas the second one is of ${\cal O}(x_{f_s}^2)$, 
hence playing a minor r\^ole. In order to extract 
$\gamma$, we may also employ $D$ decays into CP non-eigenstates 
$f_{\rm NE}$, where we have to deal with complications originating from 
$D^0,\overline{D^0}\to f_{\rm NE}$ interference effects \cite{KayLo}.
Also in this case, $\Gamma_{+-}^{f_s}$ is a very powerful ingredient, 
offering an efficient, {\it analytical} strategy to include these 
interference effects in the extraction of $\gamma$ \cite{RF-gam-det-03}.

Let us now briefly come back to the $r=d$ case, corresponding to 
$B_s\to DK_{\rm S(L)}$, $B_d\to D\pi^0, D\rho^0$ ...\ decays, which 
can be described through the same formulae as their $r=s$ counterparts. 
Since the relevant interference effects are governed by
$x_{f_d}\approx -0.02$, these channels are not as attractive for the 
extraction of $\gamma$ as the $r=s$ modes. On the other 
hand, the relation
\begin{equation}
\eta_{f_d}\langle S_{f_d}\rangle_-=
\sin\phi_q + {\cal O}(x_{f_d}^2)
=\sin\phi_q + {\cal O}(4\times 10^{-4})
\end{equation}
offers very interesting determinations of $\sin\phi_q$ \cite{RF-gam-eff-03}. 
Following this avenue, there are no penguin uncertainties, and the theoretical 
accuracy is one order of magnitude better than in the ``conventional'' 
$B_d\to J/\psi K_{\rm S}$, $B_s\to J/\psi \phi$ strategies. In particular,
$\phi_s^{\rm SM}=-2\lambda^2\eta$ could, in principle, be determined with 
a theoretical uncertainty of only ${\cal O}(1\%)$, in contrast to the
extraction from the $B_s\to J/\psi \phi$ angular distribution, which
suffers from generic penguin uncertainties at the $10\%$ level. 

Let us finally note that $\overline{B^0_d}\to D^0\pi^0$ has 
already been measured at the $B$ factories, with branching ratios at the 
$3\times 10^{-4}$ level \cite{Bbar-D0pi0}. Interestingly, the Belle 
collaboration has recently announced the observation of 
$\overline{B^0_d}\to D^0 \overline{K^0}$, with the branching ratio 
$(5.0^{+1.3}_{-1.2}\pm0.6)\times10^{-5}$ \cite{Belle-BdDK-obs}.

\subsection{$B_s\to D_s^{(\ast)\pm} K^\mp, ...$ and 
$B_d\to D^{(\ast)\pm} \pi^\mp, ...$}
Let us now consider the colour-allowed counterparts of the 
$B_q\to D f_q$ modes discussed above, which we may write generically 
as $B_q\to D_q \overline{u}_q$ \cite{RF-gam-ca}. The characteristic feature 
of these transitions is that both a $B^0_q$ and a $\overline{B^0_q}$ meson 
may decay into $D_q \overline{u}_q$, thereby leading to interference between 
$B^0_q$--$\overline{B^0_q}$ mixing and decay processes, which involve the 
weak phase $\phi_q+\gamma$ (see ii):
\begin{itemize}
\item In the case of $q=s$, i.e.\ 
$D_s\in\{D_s^+, D_s^{\ast+}, ...\}$ and $u_s\in\{K^+, K^{\ast+}, ...\}$, 
these interference effects are governed by a hadronic parameter 
$x_s e^{i\delta_s}\propto R_b\approx0.4$, and hence are large. 
\item In the case of $q=d$, i.e.\ $D_d\in\{D^+, D^{\ast+}, ...\}$ 
and $u_d\in\{\pi^+, \rho^+, ...\}$, the interference effects are described 
by $x_d e^{i\delta_d}\propto -\lambda^2R_b\approx-0.02$, and hence are tiny. 
\end{itemize}
In the following, we shall only consider $B_q\to D_q \overline{u}_q$ modes, 
where at least one of the $D_q$, $\overline{u}_q$ states is a pseudoscalar 
meson; otherwise a complicated angular analysis has to be performed
\cite{FD2}--\cite{GPW}.
 
It is well known that such decays allow a determination of
$\phi_q+\gamma$, where the ``conventional'' approach works
as follows \cite{DuSa}--\cite{BsDsK}: if we measure the observables 
$C(B_q\to D_q\overline{u}_q)\equiv C_q$ 
and $C(B_q\to \overline{D}_q u_q)\equiv \overline{C}_q$ provided by the
$\cos(\Delta M_qt)$ pieces of the time-dependent rate asymmetries, 
we may determine $x_q$ from terms entering at the $x_q^2$ 
level. In the case of $q=s$, we have $x_s={\cal O}(R_b)$, 
implying $x_s^2={\cal O}(0.16)$, so that this may actually be possible, 
though challenging. On the other hand, 
$x_d={\cal O}(-\lambda^2R_b)$ is doubly Cabibbo-suppressed. 
Although it should be possible to resolve terms of ${\cal O}(x_d)$, 
this will be impossible for the vanishingly small $x_d^2={\cal O}(0.0004)$ 
terms, so that other approaches to fix $x_d$ are required
\cite{DuSa}--\cite{SCR}. In order to extract $\phi_q+\gamma$, the 
mixing-induced observables $S(B_q\to D_q\overline{u}_q)\equiv S_q$ and 
$S(B_q\to \overline{D}_q u_q)\equiv \overline{S}_q$ associated with the
$\sin(\Delta M_qt)$ terms of the time-dependent rate 
asymmetries must be measured, where it is convenient to introduce
\begin{equation}\label{Savpm}
\langle S_q\rangle_\pm\equiv
\left(\overline{S}_q\pm S_q\right)/2.
\end{equation}
If we assume that $x_q$ is known, we may consider
\begin{eqnarray}
s_+&\equiv& (-1)^L
\left[\frac{1+x_q^2}{2 x_q}\right]\langle S_q\rangle_+
=+\cos\delta_q\sin(\phi_q+\gamma)\\
s_-&\equiv&(-1)^L
\left[\frac{1+x_q^2}{2x_q}\right]\langle S_q\rangle_-
=-\sin\delta_q\cos(\phi_q+\gamma),
\end{eqnarray}
yielding
\begin{equation}\label{conv-extr}
\sin^2(\phi_q+\gamma)=\frac{1+s_+^2-s_-^2}{2} \pm
\sqrt{\frac{(1+s_+^2-s_-^2)^2-4s_+^2}{4}},
\end{equation}
which implies an eightfold solution for $\phi_q+\gamma$. If we assume that
$\mbox{sgn}(\cos\delta_q)>0$, as suggested by factorization, a fourfold 
discrete ambiguity emerges. Note that this assumption allows us also to 
extract the sign of $\sin(\phi_q+\gamma)$ from $\langle S_q\rangle_+$. 
To this end, the factor $(-1)^L$, where $L$ is the $D_q\overline{u}_q$ 
angular momentum, has to be properly taken into account \cite{RF-gam-ca}. 
This is crucial for the extraction of the sign of 
$\sin(\phi_d+\gamma)$ from $B_d\to D^{\ast\pm}\pi^\mp$ 
modes, allowing us to distinguish between the two solutions shown in
Fig.\ \ref{fig:rho-eta-Bpipi}.

Let us now discuss new strategies to exploit the interesting physics
potential of the $B_q\to D_q \overline{u}_q$ modes, following 
\cite{RF-gam-ca}. If the width difference $\Delta\Gamma_s$ of the $B_s$ mass 
eigenstates is sizeable, the ``untagged'' rates 
\begin{eqnarray}
\lefteqn{\langle\Gamma(B_q(t)\to D_q\overline{u}_q)\rangle=
\langle\Gamma(B_q\to D_q\overline{u}_q)\rangle 
e^{-\Gamma_qt}}\label{untagged}\\
&&\times\left[\cosh(\Delta\Gamma_qt/2)-{\cal A}_{\rm \Delta\Gamma}
(B_q\to D_q\overline{u}_q)\,\sinh(\Delta\Gamma_qt/2)\right]\nonumber
\end{eqnarray}
and their CP conjugates provide observables 
${\cal A}_{\rm \Delta\Gamma}(B_s\to D_s\overline{u}_s)
\equiv {\cal A}_{\rm \Delta\Gamma_s}$ and 
${\cal A}_{\rm \Delta\Gamma}(B_s\to \overline{D}_s u_s)\equiv 
\overline{{\cal A}}_{\rm \Delta\Gamma_s}$, which yield 
\begin{equation}\label{untagged-extr}
\tan(\phi_s+\gamma)=
-\left[\frac{\langle S_s\rangle_+}{\langle{\cal A}_{\rm \Delta\Gamma_s}
\rangle_+}\right]
=+\left[\frac{\langle{\cal A}_{\rm \Delta\Gamma_s}
\rangle_-}{\langle S_s\rangle_-}\right],
\end{equation}
where the $\langle{\cal A}_{\rm \Delta\Gamma_s}\rangle_\pm$ are defined
in analogy to (\ref{Savpm}). These relations allow an 
{\it unambiguous} extraction of $\phi_s+\gamma$, if we employ again
$\mbox{sgn}(\cos\delta_q)>0$. Another important
advantage of (\ref{untagged-extr}) is that we do {\it not} have to rely on 
${\cal O}(x_s^2)$ terms, as $\langle S_s\rangle_\pm$ and 
$\langle {\cal A}_{\rm \Delta\Gamma_s}\rangle_\pm$ are proportional to $x_s$. 
On the other hand, we need a sizeable value of $\Delta\Gamma_s$. Measurements 
of untagged rates are also very useful in the case of vanishingly small 
$\Delta\Gamma_q$, since the ``unevolved'' untagged rates 
in (\ref{untagged}) offer various interesting strategies to determine 
$x_q$ from the ratio of $\langle\Gamma(B_q\to D_q\overline{u}_q)\rangle+
\langle\Gamma(B_q\to \overline{D}_q u_q)\rangle$ and CP-averaged rates of
appropriate $B^\pm$ or flavour-specific $B_q$ decays.

If we keep the hadronic parameter $x_q$ and the associated strong phase
$\delta_q$ as ``unknown'', free parameters in the expressions for the
$\langle S_q\rangle_\pm$, we obtain 
\begin{equation}
|\sin(\phi_q+\gamma)|\geq|\langle S_q\rangle_+|, \quad
|\cos(\phi_q+\gamma)|\geq|\langle S_q\rangle_-|,
\end{equation}
which can straightforwardly be converted into bounds on $\phi_q+\gamma$. 
If $x_q$ is known, stronger constraints are implied by 
\begin{equation}\label{bounds}
|\sin(\phi_q+\gamma)|\geq|s_+|, \quad
|\cos(\phi_q+\gamma)|\geq|s_-|.
\end{equation}
Once $s_+$ and $s_-$ are known, we may of course determine
$\phi_q+\gamma$ through the ``conventional'' approach, using 
(\ref{conv-extr}). However, the bounds following from (\ref{bounds})
provide essentially the same information and are much simpler to 
implement. Moreover, as discussed in detail in \cite{RF-gam-ca}
for several examples, the bounds following from $B_s$ and $B_d$ 
modes may be highly complementary, thereby providing
particularly narrow, theoretically clean ranges for $\gamma$. 

Let us now further exploit the complementarity between
$B_s^0\to D_s^{(\ast)+}K^-$ and $B_d^0\to D^{(\ast)+}\pi^-$ modes.
If we look at the corresponding decay topologies, we observe that
these channels are related to each other through an interchange of 
all down and strange quarks. Consequently, the $U$-spin symmetry 
implies $a_s=a_d$ and $\delta_s=\delta_d$, where $a_s=x_s/R_b$ and 
$a_d=-x_d/(\lambda^2 R_b)$ are the ratios of hadronic matrix elements 
entering $x_s$ and $x_d$, respectively. There are various possibilities 
to implement these relations \cite{RF-gam-ca}. A particularly simple
picture emerges if we assume that $a_s=a_d$ {\it and} $\delta_s=\delta_d$, 
which yields
\begin{equation}
\tan\gamma=-\left[\frac{\sin\phi_d-S
\sin\phi_s}{\cos\phi_d-S\cos\phi_s}
\right]\stackrel{\phi_s=0^\circ}{=}
-\left[\frac{\sin\phi_d}{\cos\phi_d-S}\right].
\end{equation}
Here we have introduced
\begin{equation}
S=-R\left[\frac{\langle S_d\rangle_+}{\langle S_s\rangle_+}\right]
\end{equation}
with
\begin{equation}
R= \left(\frac{1-\lambda^2}{\lambda^2}\right)
\left[\frac{1}{1+x_s^2}\right],
\end{equation}
where $R$ can be fixed from untagged $B_s$ rates through
\begin{equation}
R=\left(\frac{f_K}{f_\pi}\right)^2 
\frac{\Gamma(\overline{B^0_s}\to D_s^{(\ast)+}\pi^-)+
\Gamma(B^0_s\to D_s^{(\ast)-}\pi^+)}{\langle\Gamma(B_s\to D_s^{(\ast)+}K^-)
\rangle+\langle\Gamma(B_s\to D_s^{(\ast)-}K^+)\rangle}.
\end{equation}
Alternatively, we may {\it only} assume that $\delta_s=\delta_d$ {\it or} 
that $a_s=a_d$, as discussed in detail in \cite{RF-gam-ca}. 
Apart from features related to multiple discrete ambiguities, the 
most important advantage with respect to the ``conventional'' approach 
is that the experimental resolution of the $x_q^2$ terms is not required. In 
particular, $x_d$ does {\it not} have to be fixed, and $x_s$ may only enter 
through a $1+x_s^2$ correction, which can straightforwardly be determined 
through 
untagged $B_s$ rate measurements. In the most refined implementation of this 
strategy, the measurement of $x_d/x_s$ would {\it only} be interesting for 
the inclusion of $U$-spin-breaking effects in $a_d/a_s$. Moreover, we may 
obtain interesting insights into hadron dynamics and $U$-spin-breaking 
effects.

\section{Conclusions and Outlook}\label{sec:concl}
In order to perform stringent tests of the SM description of 
CP violation, it is essential to determine the angle $\gamma$ of the
UT in a variety of ways. We may divide the corresponding strategies
into methods employing the flavour symmetries of strong interactions 
and theoretically clean approaches.

As far as the former avenue is concerned, $B\to\pi K$ decays are an
important representative. Interestingly, the $B$-factory data both 
for the charged and for the neutral $B\to\pi K$ modes point separately 
towards $\gamma\gsim90^\circ$, which would be larger than the typical 
range following from the CKM fits. On the other hand, the data favour 
also conflicting ranges for the corresponding strong phases, so that we 
arrive at a puzzling picture. 
The experimental uncertainties do not yet allow us to draw definite 
conclusions, but the situation will improve in the future. A particularly 
promising strategy for $B$-decay experiments at hadron colliders is 
offered by $B_s\to K^+K^-$, which complements $B_d\to\pi^+\pi^-$ nicely, 
thereby allowing the determination of $\gamma$ with the help of a ``minimal'' 
$U$-spin input. Since $B_s\to K^+K^-$ is not accessible at the $e^+e^-$ 
$B$ factories operating at $\Upsilon(4S)$, we may not yet confront this 
strategy with data. However, we may approximately replace $B_s\to K^+K^-$ 
by $B_d\to\pi^\mp K^\pm$ to deal with the penguin effects in 
$B_d\to\pi^+\pi^-$. Interestingly, the analysis of the corresponding data 
suggests that we may accommodate $\gamma>90^\circ$ for the ``unconventional'' 
solution $\phi_d=133^\circ$ of the $B^0_d$--$\overline{B^0_d}$ mixing phase, 
whereas we obtain a picture in accordance with the SM for $\phi_d=47^\circ$.

In our discussion of theoretically clean strategies, we have focused on 
two new approaches: first, we have seen that $B_d\to D K_{\rm S(L)}$, 
$B_s\to D\eta^{(')}, D\phi$, ...\ modes provide efficient and unambiguous 
extractions of $\tan\gamma$ if we combine an ``untagged'' rate asymmetry with 
mixing-induced observables. On the other hand, the $B_s\to D_\pm K_{\rm S(L)}$,
$B_d\to D_\pm \pi^0, D_\pm \rho^0$, ...\ counterparts of these decays are 
not as attractive for the determination of $\gamma$, but allow extremely
clean extractions of $\sin\phi_s$ and $\sin\phi_d$, which may be particularly
interesting for the $\phi_s$ case. Second, we have discussed interesting new 
aspects of $B_s\to D_s^{(\ast)\pm} K^\mp$, ...\ and 
$B_d\to D^{(\ast)\pm} \pi^\mp$, ...\ decays. The observables of these modes
provide clean bounds on $\phi_q+\gamma$, where the resulting ranges for 
$\gamma$ may be highly complementary in the $B_s$ and $B_d$ cases, thereby 
yielding stringent constraints on $\gamma$. Moreover, it is of great 
advantage to combine the $B_d\to D^{(\ast)\pm} \pi^\mp$ modes with their 
$U$-spin counterparts $B_s\to D_s^{(\ast)\pm} K^\mp$, allowing us to 
overcome the main problems of the ``conventional'' strategies to deal with 
these modes. We strongly encourage detailed feasibility studies
of these new strategies.

\end{document}